\documentclass[letterpaper,11pt,superscriptaddress,nofootinbib]{revtex4-2}
\pdfoutput=1
\usepackage{xcolor}
\usepackage{amsmath,amsthm,amsfonts,amssymb,braket}
\usepackage[pdftex,colorlinks=true,linkcolor=blue,citecolor=darkred,urlcolor=darkblue]{hyperref}
\usepackage[all]{xy}
\usepackage{enumitem}	
\usepackage{fullpage}
\usepackage{mathtools}

\newtheorem{proposition-definition}[lemma]{Proposition-Definition}
\theoremstyle{definition}

\definecolor{darkblue}{rgb}{0.,0.,0.4}
\definecolor{darkred}{rgb}{0.5,0.,0.}
\definecolor{darkpurple}{rgb}{0.5,0.,0.5}
\definecolor{ltgreen}{rgb}{0.1,.59,.43}
\definecolor{orange}{rgb}{1.0, 0.5, 0.0}

\makeatletter
\def\l@subsubsection#1#2{}
\makeatother
\newcommand{\nocontentsline}[3]{}
\newcommand{\tocless}[2]{\bgroup\let\addcontentsline=\nocontentsline#1{#2}\egroup}

\newcommand{\santabarbara}{Microsoft Station Q, Santa Barbara, California, USA}

\begin{document}

\title{Fault-Tolerant Hastings--Haah Codes in the Presence of Dead Qubits}

\author{David Aasen}
\affiliation{\santabarbara}
\author{Jeongwan Haah}
\affiliation{\santabarbara}
\author{Parsa Bonderson}
\affiliation{\santabarbara}
\author{Zhenghan Wang}
\affiliation{\santabarbara}
\affiliation{Department of Mathematics, University of California, Santa Barbara, USA}
\author{Matthew Hastings}
\affiliation{\santabarbara}

\begin{abstract}
We develop protocols for Hastings--Haah Floquet codes in the presence of dead qubits.
\end{abstract}

\date{\today}
\maketitle


\section{Introduction}

A fundamental problem in building a quantum computer is dealing with imperfect components.
One aspect of this problem is that any operation on physical qubits, whether it be a gate, a measurement, or even simply sitting idle, will introduce some noise into the system.
To deal with this, quantum error correction will be necessary~\cite{Shor1997faulttolerant}.
However, another aspect of the problem is that some fraction of the qubits may have manufacturing defects that result in those qubits being ``dead,'' either being completely nonfunctional or having a noise rate significantly higher than most other qubits.
This issue is particularly acute in hardware architectures where qubits are arranged in a planar geometry with local couplings between qubits.
In this case, dead qubits introduce locations where logical operators can terminate, which can substantially reduce the code distance and performance.
One might think that we could just run the error correcting code as usual and account for dead qubits in
classical post-processing, but that is unlikely to be true.

Let us consider in more detail why dead qubits are a problem for planar codes, such as the surface code~\cite{Kitaev2003,Bravyi1998} or the Hastings--Haah (HH) Floquet code~\cite{hastings2021dynamically,Haah2022}.
If we do not modify the measurement schedule in some way, then each dead qubit acts as a puncture in the planar code, which results in an extra logical qubit with low weight logical operators supported near the puncture.
Suppose we have some finite density of dead qubits with a typical seperation distance $\ell$ between them.  
With some probability that is exponentially suppressed in $\ell$, noise in the system can create an undetectable local logical operator whose end points are localized to a pair of dead qubits a distance $\ell$ apart.
The probability of this occurrence is exponentially small in $\ell$, but it is not suppressed in system size.
Without modification, running the code does not measure the local logical operators encircling the dead qubits and, hence, does not detect such error processes.
As we continue running the code, at a later time, another logical operator can be created that terminates on a different pair of dead qubits, and so on until the system is full of undetectable errors, and error correction is rendered impossible.  
Indeed, the code will no longer have a nonzero threshold against noise in this case.
Thus, it is essential to measure the logical operators encircling the defects. 
For the surface code, strategies for dealing with dead qubits were proposed in~\cite{Stace2009,Stace2010,Tang2016,Auger2017,Nagayama2017,Strikis2021,Siegel2022}, where dead qubits are alternatively referred to as ``loss,'' ``fabrication errors,'' ``defects,'' or ``vacancies'' and the resulting local logical operators are called ``superstabilizers''.

In this paper, we propose several strategies for dealing with dead qubits in the context of Hastings--Haah (HH) Floquet codes~\cite{hastings2021dynamically,Haah2022}.
One of our solutions -- recoupling the lattice -- can be applied to any code running on a plaquette 3-colorable lattice, see Sec.~\ref{sec:lattice_recouple}.
This solution can simply be understood as a re-triangulation of the 3-colorable lattice so that none of the dead qubits are used in the resulting measurement schedule.
We primarily restrict our attention to a single faulty qubit, and design a measurement sequence which does not have support on the dead qubit.
We discuss how this can be generalized to deal with a larger number of dead qubits by applying the method iteratively.~\footnote{We note that these techniques may also find application in scenarios where one deliberately introduces logical qubits into a code by adding punctures to the surface.
In such situations, our techniques could be adapted to provide a way to measure logical operators of the logical qubits associated with the punctures.
Similarly, they could be used to measure joint logical operators of several logical qubits defined by punctures.}
We also discuss issues of optimizing how the strategy is applied, including considerations specific to implementations in Majorana-based hardware.
In Sec.~\ref{sec:decouple}, we propose another solution for dealing with dead qubits -- decoupling with $1$-gon measurements.
In this solution, we essentially introduce a boundary to the code lattice where we remove plaquettes, but specifically order the measurements such that the local logical operator, i.e., the superplaquette operator, is measured.
Again this technique can remove any plaquette in the system, and can be applied iteratively until all dead qubits have been removed from the measurement sequence.
Finally, we propose a solution -- the triangle sequence -- which can only account for one isolated dead qubit. 
The benefit of this method is that it does not exclude functioning qubits from the modified code.
In Sec.~\ref{sec:identifying_dead_qubits}, we discuss the identification of dead qubits, possibly during computations.
Finally, in Sec.~\ref{sec:threshold}, we provide an argument that the recoupling and decoupling solutions result in a threshold in the thermodynamic limit for sufficiently small error rates and sufficiently low density of dead qubits.

\section{Dead Qubit Strategies for the Hastings--Haah Code}

We begin with a rapid review of the HH code.
The HH code can run on any plaquette 3-colorable lattice, in which case qubits are located at the vertices of the lattice and the edges connecting pairs of qubits indicate which two-qubit measurements are required for code operation.
A plaquette 3-colorable lattice is a trivalent lattice where each plaquette can be colored one of three colors (we will use red, blue, and green), such that no two adjacent plaquettes have the same color.
The plaquette coloring induces an edge coloring, where the color of an edge is different from the two plaquettes it borders, i.e. an edge is red if it is shared by a blue and a green plaquette.
The HH code consists of a sequence of pairwise measurements. 
Denote the set of pairwise $XX$ measurements on qubits connected by red edges as $E_r(X)$, the set of pairwise $YY$ measurements on qubits connected by green edges as $E_g(Y)$, and pairwise $ZZ$ measurements on qubits connected by blue edges as $E_b(Z)$.
The HH code without boundary is given by the following period three sequence of measurements~\cite{hastings2021dynamically}
\begin{align}
   \cdots \rightarrow E_r(X) \rightarrow E_g(Y) \rightarrow E_b(Z) \rightarrow \cdots
.
\end{align}
In a planar geometry with a boundary, it is helpful to ``unwind'' the automorphism with the following period six sequence~\cite{Haah2022}
\begin{align}
\label{eq:period6}
\cdots \rightarrow E_r(X) \rightarrow E_g(Y) \rightarrow E_b(Z) \rightarrow E_r(X) \rightarrow E_b(Z) \rightarrow E_g(Y) \rightarrow \cdots
. 
\end{align}
The HH code has static plaquette stabilizers given by a product of the measurements around any given plaquette.
All plaquette stabilizers of color $c$ are inferred by measuring the edge operators of two distinct colors, i.e., $E_{c'}$ and $E_{c''}$ where $c'$ and $c''$ are distinct from $c$.
This inference of plaquette stabilizers, or more precisely the change of its value in time,
defines \emph{detectors} by which we perform error correction for each of the two realizations of the HH code written above.
Here, a detector is a bit of information 
that is always zero in the absence of any faults
and that we use in decoding algorithms.

\subsection{Recoupling the lattice}
\label{sec:lattice_recouple}

The conceptually simplest solution for dealing with dead qubits in the HH code is to ``recouple'' the code lattice in the vicinity of dead qubits in a manner that yields another plaquette 3-colorable lattice which excludes the dead qubits.
Recoupling the code lattice in this context means modifying the set of two-qubit measurements that are utilized in the code operation.
Since the recoupled lattice is, by design, plaquette 3-colorable, we can simply run the HH code on it instead of the original lattice.
This solution effectively patches the code over the dead qubits.
We remark that this technique of dealing with dead qubits can be applied to any quantum error correcting code that runs on a plaquette 3-colorable lattice, which includes the HH code and the color code as examples.
Indeed, from a more general topological perspective, this recoupling strategy is a re-triangulation of the code lattice, which is a general strategy that can, in principle, be applied to many topological quantum error correcting codes, irrespective of 3-colorability.
While conceptually simple, the key challenge in this strategy is to find recouplings for which the set of measurements involved in the recoupled code lattice are physically realistic to implement.

The primitive operation of our recoupling strategy for dead qubits in the HH code is to remove a full plaquette and all the qubits located along its boundary.
In order to remove a plaquette, we first remove all qubits and edges along the boundary of that plaquette.
This will result in an even number of dangling edges, i.e. each edge that connected a removed qubit to one that is not being removed.
We then pair up these dangling edges, rerouting the connections between qubits that were formerly connected to removed qubits.
There are two possible ways to recouple the dangling edges: labeling the $n$ dangling edges with integers $\{ 1,\cdots, n\}$ (where $n$ is even), we can either recouple $2j-1$ to $2j$ (mod $n$) for all edges or $2j$ to $2j+1$ (mod $n$) for all edges.
This can be applied to any plaquette in a plaquette 3-colorable lattice, and the resulting recoupled lattice will still be plaquette 3-colorable.

One strategy for dealing with multiple dead qubits is to apply the recoupling primitive (plaquette removal) iteratively, until we have removed all the dead qubits.
There is typically freedom to choose which plaquettes to remove in this process, and these choices should be optimized.
With this strategy, two natural optimizations are: (1) minimizing the number of properly functioning qubits that are excluded from the recoupled code lattice, i.e. efficiently utilizing the good components, which we expect helps maximize the code distance; and (2) minimizing the number of qubits (or edges) along the boundary of the plaquettes in the recoupled lattice, which we expect helps performance since detectors have a fidelity that drops with the number of contributing measurements.
These optimizations may be in contention with each other, depending on the lattice realization; that is, minimizing the number of functioning qubits that are excluded may not minimize the number of contributing measurements to a given detector.
Moreover, one must consider optimizations with respect to the particular hardware implementation being utilized (and these may also be in contention with the previous optimizations).
For example, if a choice of recoupling results in measurements which are not native to the hardware or are prohibitively difficult to perform with desired fidelity, that recoupling may not represent a suitable way of dealing with the dead qubits.

\begin{figure}[t!]
\centering
\includegraphics[width=.95\linewidth]{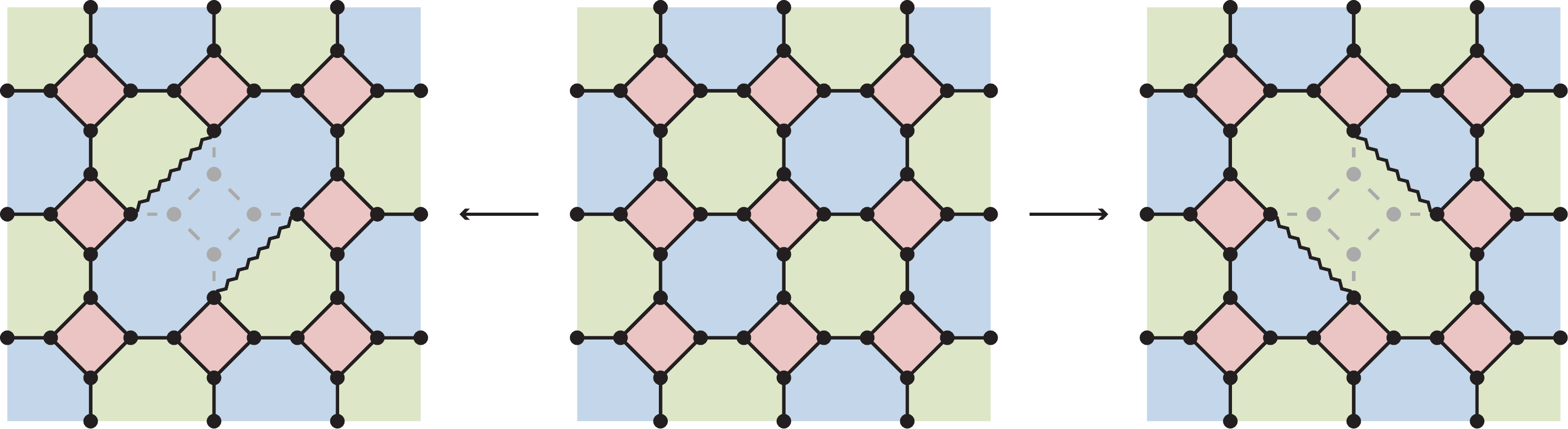}
\caption{The recoupling primitive applied to a $4$-gon on the 4.8.8 lattice.
Black dots represent qubits in the code and grey dots represent qubits that have been excluded by the recoupling.
(The induced coloring of edges is implied by the plaquette coloring, but left implicit in this figure.)
The operation removes one entire plaquette and all the qubits along its boundary, in this case the red plaquette at the center.
On the left, the dangling (red) edges are recoupled across the green plaquettes; on the right, the dangling (red) edges are recoupled across the blue plaquettes.
The new (red) edges are shown as zig-zag lines and the removed edges are shown as dashed lines.
These new zig-zag lines correspond to the introduction of new measurements in the code operation schedule.
In this case, both recouplings result in a larger 12 qubit plaquette (blue on the left and green on the right), while two of the octogons are reduced to hexagons (green on the left and blue on the right).
For the 4.8.8 lattice, any single qubit will be located along a $4$-gon, and hence can be excluded by a recoupling that removes one $4$-gon.
}
\label{fig:plaquette_removal}
\end{figure}
\begin{figure}[t!]
\centering
\includegraphics[width=.95\linewidth]{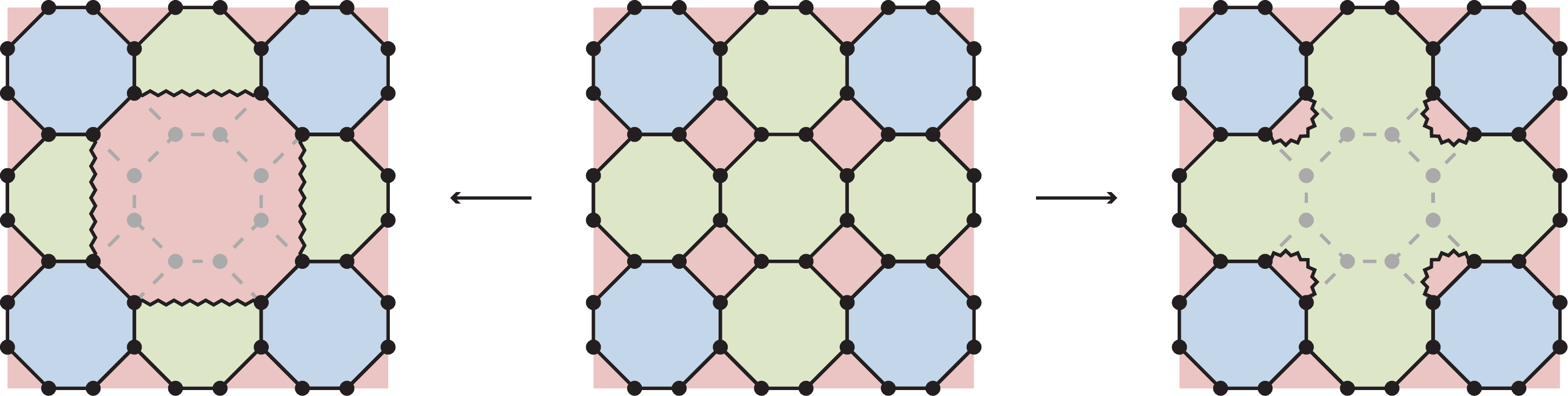}
\caption{The recoupling primitive applied to an $8$-gon on the 4.8.8 lattice.
In this case, the two recoupling options result in significantly different recoupled lattices and resulting detectors, so these two options are not expected to have equivalent performance.
}
\label{fig:plaquette_removal_prime}
\end{figure}

We now consider specific examples of these issues in the context of the 4.8.8 HH code realized in Majorana-based hardware.
(Similar analysis can be performed for the HH code on other lattice realizations and other hardware implementations.)
We begin by considering an abstract 4.8.8 lattice before including hardware considerations.
We demonstrate the recoupling primitive applied to the HH code on the 4.8.8 lattice in Figs.~\ref{fig:plaquette_removal} and \ref{fig:plaquette_removal_prime}, showing the two possible recouplings in both cases.

Notice that any single qubit is always in the support of one $4$-gon and two 8-gons.
If only a single qubit in the region is dead, then clearly the recouplings in Fig.~\ref{fig:plaquette_removal} are more efficient with its exclusion of functioning qubits than the recoupling in Fig.~\ref{fig:plaquette_removal_prime} (three vs. seven functioning qubits excluded).
If multiple nearby qubits in a small region are dead, the accounting can change, i.e. there are configurations of dead qubits that may be excluded by removing a single octagonal plaquette which could not be excluded by removing two $4$-gons.

In Fig.~\ref{fig:plaquette_removal}, the two recoupling options are symmetric, so if all other factors are equal, they would result in the same performance.
In contrast, the two recoupling options in Fig.~\ref{fig:plaquette_removal_prime} differ significantly.
We expect the option on the left to perform better than the one on the right, because of the 24-qubit plaquette that is formed on the right.

\begin{figure}[t!]
\centering
\includegraphics[width=.8\linewidth]{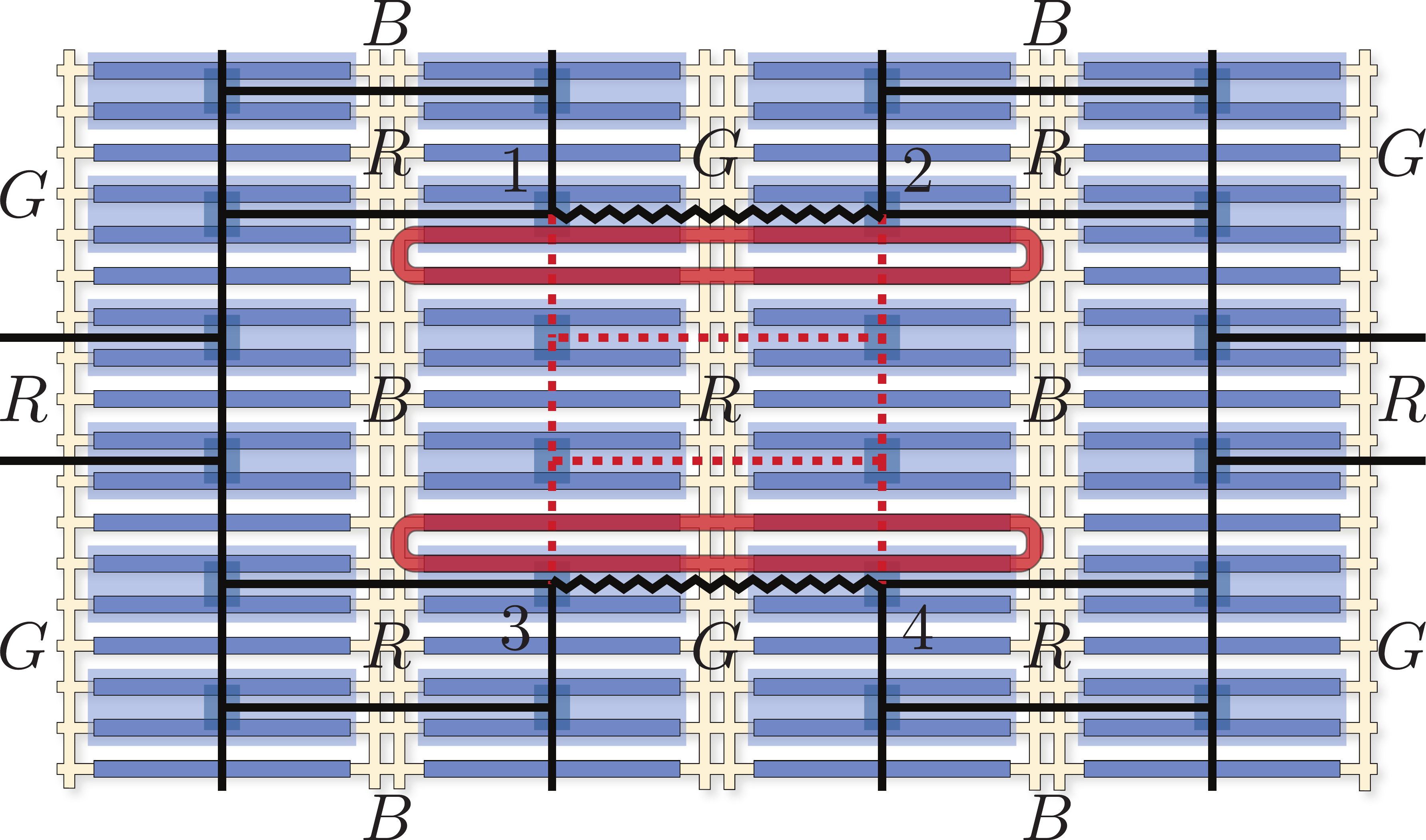}
\caption{A layout of the 4.8.8 HH code using Majorana-based hardware.
The hardware array involves tetron qubits (pairs of Majorana wires connected by a superconducting backbone), coherent links (single Majorana wires), and double rails of semiconductors that can be used to form gate-defined quantum dots to connect the zero modes that exist at the ends of the Majorana wires.
The code lattice is drawn overlaying the physical hardware array.
Here, we show the recoupling for a dead qubit in which we remove a (red) $4$-gon from the code lattice.
The qubits involved in the recoupling are labeled $1-4$. 
The zig-zag lines represent the new edges for one choice of the recoupled lattice, where qubits 1 and 2 are connected and qubits 3 and 4 are connected.
The dashed lines represent the excluded edges of the original lattice.
The new two-qubit measurements corresponding to the zig-zag lines will involve interference loops (shown as thick red lines) that necessarily require the use of two coherent links.
Alternatively, one could have chosen to recouple the lattice such that qubits 1 and 3 are connected by a new edge, and qubits 2 and 4 are connected.
Such a recoupling would involve interference loops with long distances through the semiconductor rails.
}
\label{fig:foureighteightlayout}
\end{figure}

Next, we examine the recoupling operations for a particular Majorana-based hardware implementation of the 4.8.8 HH code. 
We present a layout proposed in Ref.~\cite{Paetznick2023} of the 4.8.8 code using Majorana tetron architectures in Fig.~\ref{fig:foureighteightlayout}.
The removal of the $4$-gon results in two new measurements.
For one of the recoupling options (corresponding to the left option in Fig.~\ref{fig:plaquette_removal}), these new measurements are between horizontally adjacent (nearest neighbor) pairs of tetron qubits.
As shown in Fig.~\ref{fig:foureighteightlayout}, these new measurements require the use of two coherent links.
(Recall that the original bulk measurements of the 4.8.8 HH code involved no coherent links for the idle code operation.)
In contrast, the other recoupling option for removing a $4$-gon (corresponding to the right option in Fig.~\ref{fig:plaquette_removal}) would result in new measurements between 3rd nearest neighbor horizontally separated pairs of tetron qubits.
Such measurements would involve rather long distances through the semiconductor rails, i.e. large quantum dots, which might even include faulty components as they run adjacent to at least one dead qubit.
Thus, the two options for recoupling that seemed equivalent for the abstract system in Fig.~\ref{fig:plaquette_removal} are seen to be inequivalent when hardware considerations are included.

\subsection{Decoupling with $1$-gon measurements}
\label{sec:decouple}

We now introduce another method of dealing with dead qubits by removing plaquettes that contain them.
This method is inspired by the boundary condition introduced in~\cite{Gidney2022}.
For this solution, rather than recoupling the code lattice to patch over the excluded qubits as described in Sec.~\ref{sec:lattice_recouple}, we instead decouple the code lattice from the excluded qubits and treat the exclusion like the creation of a boundary in the code lattice and measure the resulting local logical qubits that such boundaries define.
In doing so, we avoid introducing new 2-qubit measurements, which are potentially difficult or problematic.
We instead utilize single qubit (``$1$-gon'') measurements, which are relatively simple and already in the required set of measurements for operating the device, e.g. for performing logical operations.
When the HH code is being run on a patch with a boundary, we typically use the period 6 bulk measurement sequence shown in Eq.~\eqref{eq:period6}.
Hence, it is convenient to have a measurement sequence which is commensurate with this cycle.
This approach has the advantage of requiring only a very minimal change to how we operate the hardware.

In the following, we describe the decoupling primitive of removing a single plaquette.
As with the recoupling strategy, additional plaquettes can be removed by iterating this procedure.
We remark that with the decoupling strategy, the boundary of every region of removed plaquettes must have only two-colors of plaquettes present.
Similarly, choices of which plaquettes to remove should be optimized.
We emphasize that once we specify which plaquettes are to be removed, there are no further choices to make in the decoupling approach, in contrast with the recoupling strategy.

\begin{figure}[t!]
\centering
\includegraphics[width=.97\linewidth]{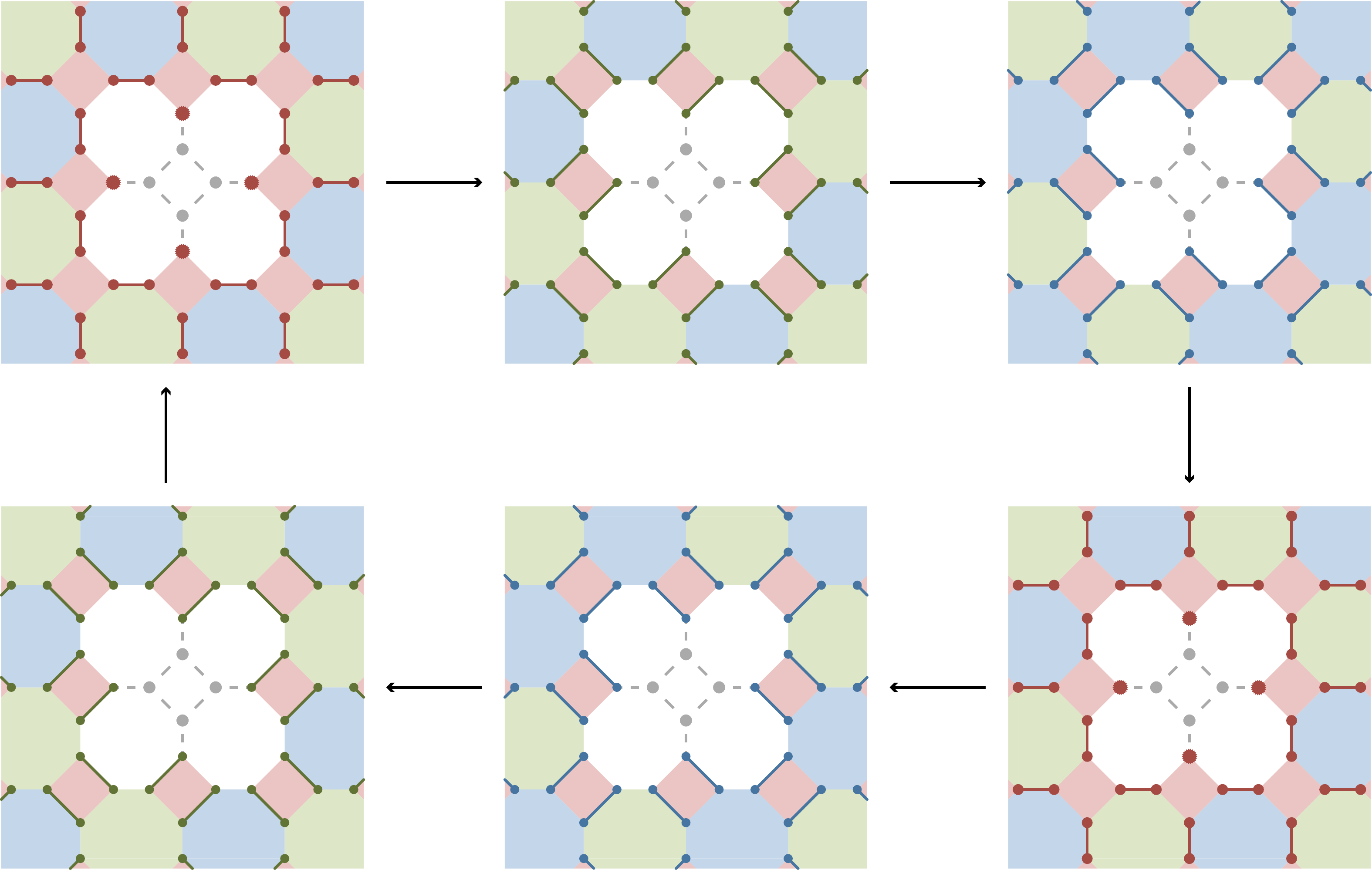}
\caption{The sequence of measurement steps of Eq.~\eqref{eq:decouple} used in an application of the decoupling primitive to remove one red $4$-gon.
The measurement sequence generates the superplaquette operator associated with the new boundary of the decoupled code lattice.
The red edges correspond to pairwise $XX$ measurements and red spiky dots correspond to single qubit $X$ measurements, associated with the new $1$-gons. 
The green edges correspond to pairwise $YY$ measurements and the blue edges to pairwise $ZZ$ measurements.}
\label{fig:decouple}
\end{figure}

As before, we remove a plaquette by excluding all the qubits along the boundary of the plaquette, and this results in an even number of dangling edges in the lattice, all of the same type.
In this solution, we modify the code such that these dangling edges are treated as $1$-gons, corresponding to single qubit measurements in the modified measurement schedule.
We call the resulting code lattice the ``decoupled'' lattice.
Let us assume we have removed a red plaquette, so that the dangling edges are all red.
We denote the modified set of red measurements $\widetilde{E}_r(X)$ for the decoupled lattice to be all the (non-excluded) pairwise measurements corresponding to red edges, together with all the single qubit measurements associated with the new $1$-gons.
As before, we denote the set of pairwise measurements corresponding to green edges as $E_g(Y)$, and to blue edges as $E_b(Z)$.
(Note that since the red plaquette has been removed, it is inconsequential whether we include the measurements that only have support on that plaquette under usual operation, as it will be decoupled from the rest of the system.)
The modified period six measurement schedule for the decoupled lattice is then given by 
\begin{align}
\label{eq:decouple}
\cdots \rightarrow \widetilde{E}_r(X) \rightarrow E_g(Y)  \rightarrow E_b(Z) \rightarrow \widetilde{E}_r(X) \rightarrow  E_b(Z)
\rightarrow 
E_g(Y) \rightarrow \cdots
.
\end{align}
This modified sequence will generate the superplaquette operator for the boundary created in the decoupled lattice.

\begin{figure}[t!]
\centering
\includegraphics[width=.8\linewidth]{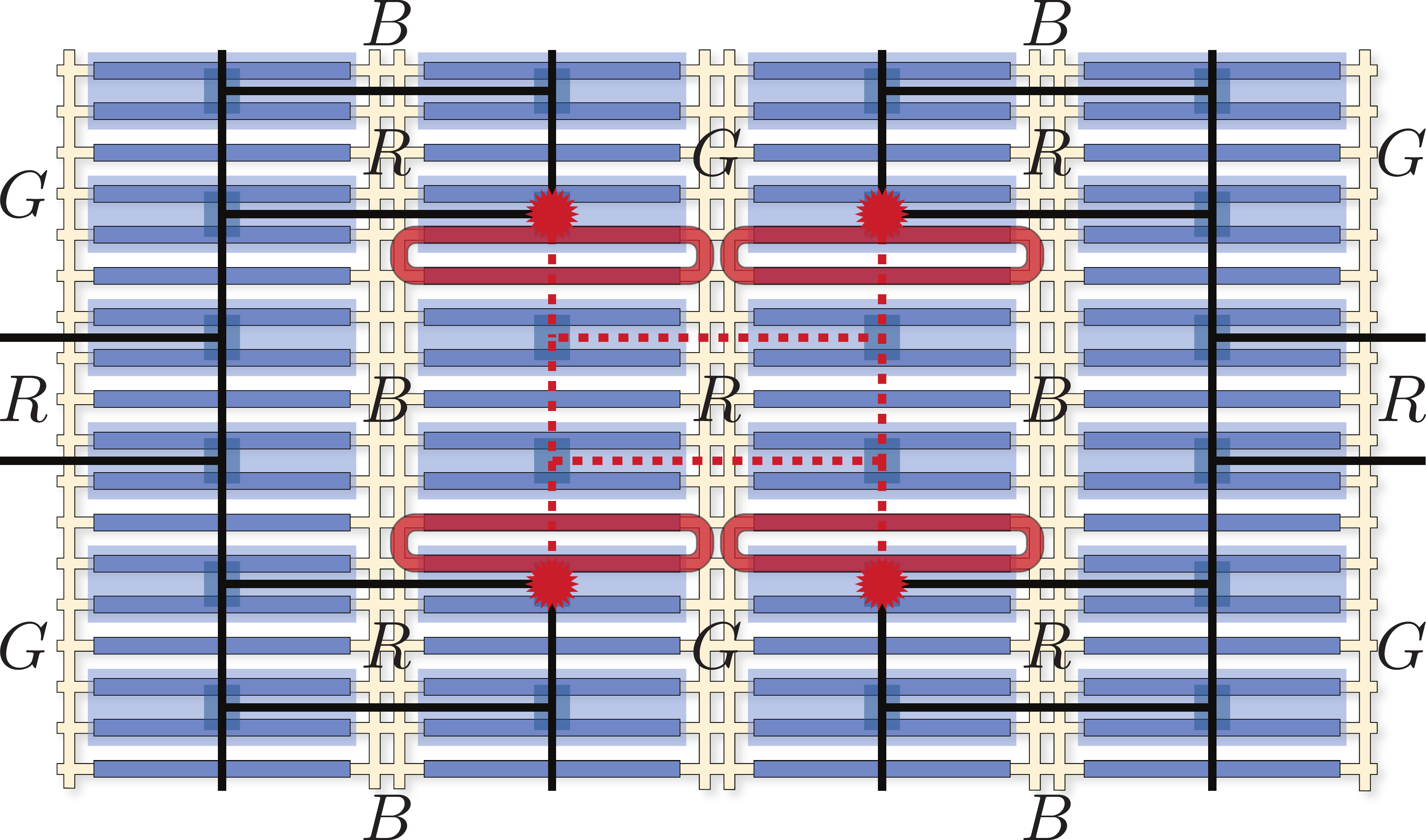}
\caption{Removal of a $4$-gon in the 4.8.8 HH code on Majorana-based hardware.
Using the decoupling strategy, the single qubit measurements are introduced into the measurement schedule for the qubits that have dangling edges as a result of removing the $4$-gon plaquette.
These single qubit $1$-gon measurements are indicated with spiky dots.
The corresponding interference loops associated with these measurements are shown by thick red loops.
}
\label{fig:physical_decouple}
\end{figure}

In Fig.~\ref{fig:decouple}, we provide an explicit example of the operation for the 4.8.8 HH code when we remove a single $4$-gon plaquette using the decoupling strategy described here.
In Fig.~\ref{fig:physical_decouple}, we examine the decoupling operations for the 4.8.8 HH code implemented in the Majorana-based hardware.
Since no new edges are introduced into the modified lattice and no new measurements added to the set of necessary operations, there are no concerns about whether new measurements are difficult or problematic.

\subsection{Triangle sequence}\label{sec:effective_stabilizers}
 
The code lattice recoupling and decoupling ideas discussed in Sec.~\ref{sec:lattice_recouple} and~\ref{sec:decouple} both involve the likely exclusion of properly functioning qubits in the process of excluding dead qubits.
Here, we provide another solution which makes use of all functioning qubits, yet also combines ideas from both Sec.~\ref{sec:lattice_recouple} and Sec.~\ref{sec:decouple} in a nontrivial manner.
This results in a deformation of the HH code with a modified measurement sequence near the dead qubit. 
It also only requires introducing two new measurements.

\begin{figure}[t!]
\centering
\includegraphics[width=.97\linewidth]{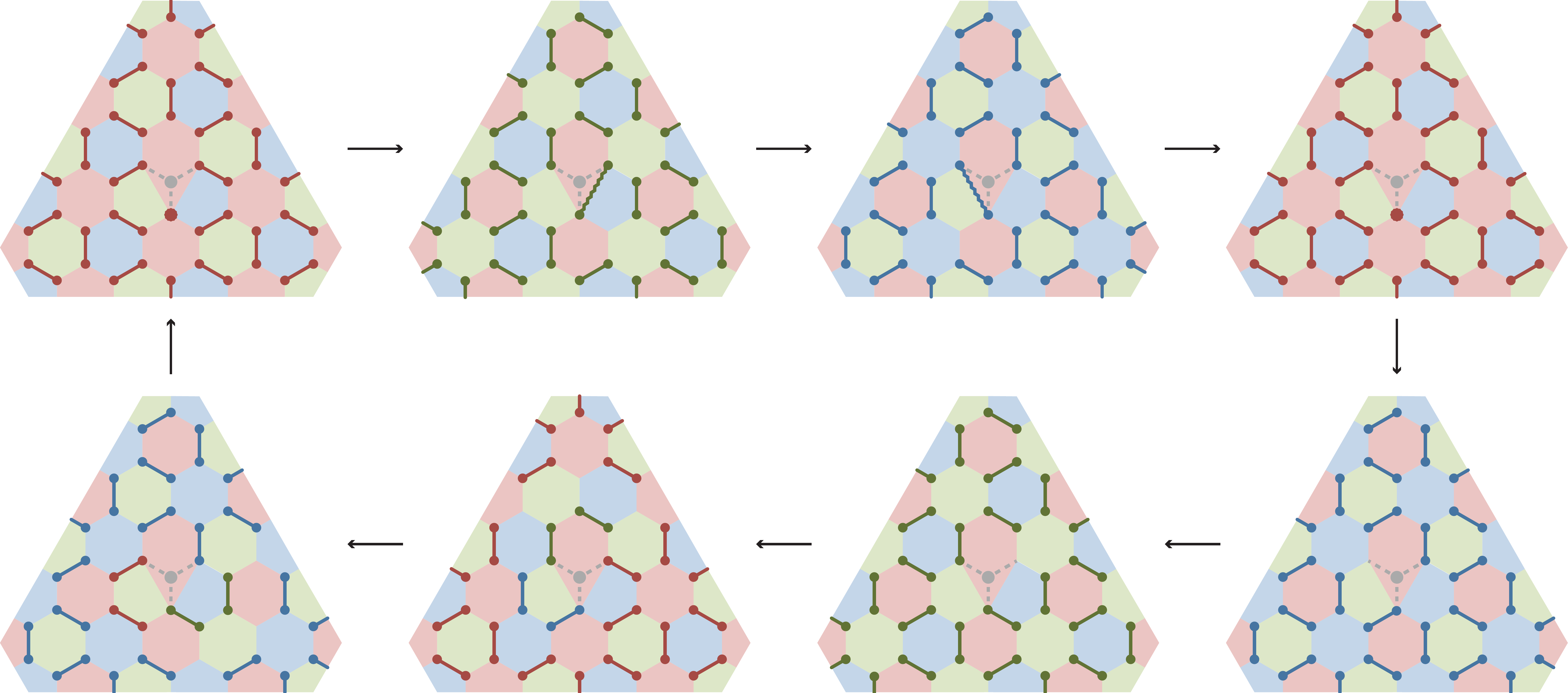}
\caption{A sequence of eight measurement steps which infers all superlattice stabilizers once per eight measurement steps.
Stabilizers far from the dead qubit are operating as in the usual honeycomb code according to 
$\cdots \rightarrow r \rightarrow g\rightarrow b\rightarrow r\rightarrow b\rightarrow g\rightarrow r\rightarrow b \rightarrow \cdots.$
Near the dead qubit we modify the measurements to include two ``new measurements'' (steps 2 and 3 from top left), one $1$-gon measurement (steps 1 and 4 from top left), as well as a schedule change from the usual $r,g,b$ measurements (last two steps, bottom left).}
\label{fig:triangle_schedule}
\end{figure}

In terms of the effective surface code, the dead qubit acts like a missing edge. 
Therefore, on say, the hexagonal lattice, the effective surface code will have a 5-valent vertex term, and a 4-sided plaquette term at the location of the missing edge at each measurement step. 
These defect vertex and plaquette terms will rotate around the dead qubit from round to round.
Our strategy here is to make a designer measurement schedule which reads out the value of the resulting effective surface code stabilizers once per measurement period.
The modified measurement sequence is presented in Fig.~\ref{fig:triangle_schedule}.

While this strategy has the advantage of being efficient with the exclusion of qubits, a disadvantage is that the measurement schedule has a longer period as some of the detectors with support near the dead qubit require a longer measurement period.
We remark that the bulk measurement sequence in Fig.~\ref{fig:triangle_schedule} may need to be modified for a planar realization, which may be necessary in practice.

\section{Identifying Dead Qubits}
\label{sec:identifying_dead_qubits}

Our strategies for dealing with dead qubits assume that we know which qubits are dead and can accordingly modify the system's operation schedule to counteract their presence.
The simplest situation is when dead qubits are identified in the process of bring up and calibration.
In this case, we modify the measurement schedule before running the code and any computation.

However, if we do not know which qubits are dead from the onset or if they die partway through the computation, their presence can be detected by running the error correcting code on the device and then applying classical post-processing to locate the problematic qubits.
The measurement outcomes for all pairwise measurements at the boundary of any given plaquette in the HH code are highly correlated.
In particular, combinations of measurement outcomes form detectors whose value remain invariant in the absence of errors.
For example, a detector in the HH code on the hexagon lattice consists of the mod~$2$ value of twelve measurement outcomes. 
When a dead qubit is involved in a detector, each round will likely result in a random result for that detector.
We can gather statistics on all detectors over many rounds and assume unreliable detectors contain one or more dead qubits in their support.
Three neighboring detectors triangulate a single dead qubit, four neighboring detectors triangulate a pair of dead qubits, and so on.
Once qubits are diagnosed in this way to be dead, we can modify the measurement schedule accordingly to exclude them from the code at the earliest possible opportunity.~\footnote{In principle, we can also reverse this process and reincorporate qubits that cease to be dead.
This would involve a protocol of monitoring dead qubits after they are excluded from the code, e.g. by performing single qubit measurements on the dead qubits to characterize their errors.
We assume the probability of qubit resurrection is very low without divine intervention, so we do not focus on this matter in detail.}

\section{Threshold}
\label{sec:threshold}

An interesting academic question is whether these techniques lead to a threshold. 
That is, suppose each qubit is dead with some given probability (independently of the other qubits), and suppose there is some circuit level noise; then, if both the probability of a dead qubit and the noise level are sufficiently small but nonzero, is the logical error probability vanishing in the thermodynamic limit? 

The question has been studied for the surface code. 
In that context, a numerical study~\cite{Auger2017} provided evidence for a threshold. 
The method of handling dead qubits in that study was, roughly, as follows: given a set of dead qubits near each other, one removes them from the lattice, giving some ``hole'' in the lattice.
The hole is given some particular boundary conditions (chosen either electric or magnetic, to use the field theory terminology, or rough or smooth to use the error correction terminology) by an appropriate choice of stabilizer to measure at the boundary.
Then, some number $O(1)$ of rounds later, the boundary conditions are changed, with the boundary conditions alternating between electric and magnetic every $O(1)$ rounds.  
Heuristically, such a hole may trap an anyon (either electric or magnetic, or their product which is a fermion), but a given choice of boundary conditions allows one to measure the presence of a given type of anyon in the hole, so by alternating boundary conditions, both types may be detected.

While no threshold has been \emph{proven} for that protocol, a threshold has been proven for a related protocol~\cite{Strikis2021}, at least for a phenomenological noise model.
In that paper, the alternation between boundary conditions was modified to follow a different schedule, so that for a hole of boundary size $r$, one would spend roughly $r$ rounds with electric boundary conditions, then roughly $r$ rounds with magnetic, and then repeating this period $2r$ cycle.

In order to address whether a threshold exists for our decoupling and recoupling dead qubit strategies, we argue that these two proposals in this paper are both close analogues to the case of the surface code where the boundary condition alternates every $O(1)$ rounds.
As such, while no noise threshold is proven, the numerical evidence of Ref.~\cite{Auger2017} may also provide evidence for our methods.

Let us explain our arguments for the analogy in two ways.
The first argument is purely heuristic, while the second argument relies on decoding graphs.
For the heuristic argument, let us recall how boundaries were introduced in Ref.~\cite{Haah2022}.
It was shown that for various boundary condition choices (including the $2$-gon and $4$-gon boundaries), one could define an effective superlattice surface code.
The boundary conditions for this code depended on which boundary detectors were measured most recently.
The goal was to introduce boundaries \emph{without} measuring the logical operators created by the boundary, in order to introduce logical qubits into a planar code.
To do this, a particular schedule for the bulk and boundary was introduced so that the effective boundary conditions alternated between electric and magnetic every round.
Recall, however, that effectively the bulk of the HH code is \emph{also alternating} every round; that is, an electric string operator in the bulk turns into a magnetic string, and vice-versa, each round.
So, if one has a particular choice of boundary conditions so that a particular type of logical operator can terminate on the boundary, in the next round, the logical operator type and boundary conditions both change so that the image of that logical operator under the automorphism can terminate on the same boundary.
One may say that
``relative to the bulk,'' the boundary is not alternating in the scheme of that paper.
In the present paper, we do something different around the boundaries of holes.
Our choice of boundary schedule means that the boundary conditions of the effective code \emph{do not} alternate every round; there is some alternation, but it is not every round.
So, ``relative to the bulk,'' the effect is as if the boundary condition is alternating every $O(1)$ rounds.

Let us also give a more formal argument for the analogy between the methods we have proposed and the surface code with boundary conditions alternating every $O(1)$ rounds.
This relies on the decoding graph for the code, which is a graph where errors may occur on edges and detectors are vertices.
Thus, a decoding graph is an appropriate description of a code where each error may flip exactly two detectors.
In the surface code, with $Z$-type plaquette operators and $X$-type vertex operators for definiteness, one may define two decoding graphs: one for $X$-type errors and the other for $Z$-type errors.
Note, however, that if one has a $Y$ error, this can flip four detectors, two on each graph.
So, defining the decoding graph requires picking a particular basis in which to expand Pauli errors.
A similar decoding graph may also be defined for the HH code~\cite{hastings2021dynamically,Haah2022}, but it requires a different choice of basis for errors, with the basis depending on the spacetime location of the qubit.
Furthermore, such a decoding graph may still be defined for the HH codes even with the decoupling or recoupling solutions proposed here; it simply corresponds to an HH code on a different lattice.

We emphasize that for the HH code using the solutions here, there are no errors that flip only a single detector.
If one instead used the boundary conditions of Ref.~\cite{Haah2022} to introduce logical qubits, then there would be such errors (indeed, such errors must exist, as there are logical operators which can terminate on the boundary without violating any detectors).

Having introduced the decoding graph, we can see the issue with proving a threshold.
The decoding graph will have high degree vertices if there are large holes; the vertex is associated with a flip in the value of the stabilizer around this hole.
Indeed, in the thermodynamic limit, the degree of the vertices is unbounded
since there will be an arbitrarily large cluster of dead qubits 
somewhere in the infinite lattice.

For the dead qubit strategies employed for the surface code in Ref.~\cite{Auger2017},
similar high degree vertices in the decoding graph also arise.
Suppose one has a given type of boundary conditions so that a given type of defect string can terminate on the boundary undetected.
For example, suppose an electric string can terminate undetected on the boundary.
When one changes to the other type of boundary conditions, one can detect that this has occurred.
More precisely, one can determine the parity of the number of strings that terminated on the boundary, but one cannot determine where in spacetime this occurred.

These high degree vertices form columns in spacetime and they prevent one from using a simple Peierls-type argument to prove a threshold, though with the schedule of Ref.~\cite{Strikis2021} it was possible to prove a threshold because high degree vertices become sufficiently separated from each other in spacetime.
We remark that we could implement something similar to the schedule of Ref.~\cite{Strikis2021} for our dead qubit strategies.
In order to do this, we keep the boundary conditions fixed ``relative to the bulk'' (i.e., strictly alternating from round to round, which can be accomplished by following the schedule of Ref.~\cite{Haah2022}) for roughly $r$ rounds, then switching them ``relative to the bulk" and then again keeping it fixed for roughly $r$ rounds, and repeating this cycle.

\begin{acknowledgments}
We thank A. Paetznick and N. Delfosse for useful discussions.
\end{acknowledgments}

\bibliographystyle{apsrev4-2}
\bibliography{references}

\end{document}